# Meta model application for consistency management of models for avionic systems design

Jef Stegen
Stefan Dutre

Siemens Industry Software
Leuven, Belgium
jef.stegen@siemens.com
stefan.dutre@siemens.com

Joe Zhensheng Guo
Marc Zeller
Stefan Rothbauer

Siemens AG, Corporate Technology
Munich, Germany
joe.guo@siemens.com
marc.zeller@siemens.com
stefan.rothbauer@siemens.com

*Abstract—* **This paper presents the application of a meta model and single underlying model on an applied avionics system design use case. System models, safety assurance cases and safety requirements are maintained in a central repository. This enables to link these data which are originally developed in unrelated tools. By having such a central repository, traceability can be established, and consistency can be ensured, which leads to less errors and a shorter development time. A meta model was constructed which matches the central repository to enable bidirectional synchronization with an external authoring tool.**

## I. Introduction

The design of new avionic systems has never been as complex as it is today. Moreover, the development must follow regulations, such as defined in the standards ARP4754 [1] and ARP4761 [2], to ensure safe operation. An avionic system often consists of numerous subsystems. These subsystems are physically distributed and include mechanical, electrical and software components. While in the past an aircraft could be considered as predominantly a mechanical system, it recently has transformed to a mechatronic system involving electronics, mechanics and software.

This increased complexity leads to more inconsistencies when subsystems are integrated. Each subsystem has its own complex characteristics and is therefore developed in specialized departments by domain specialists with domain-specific tools. During development, assumptions are often made concerning environmental conditions and operation of other parts of the system.

These inconsistencies can lead to integration issues which are only discovered late in the development process, as the integration phase often occurs when all subcomponents are finished.

Integration problems found late in the development process result in large correction costs. Subsystems are often only integrated when a prototype is being built. If a fundamental issue occurs at this stage, a large part of the system must be redesigned to solve the issue. This leads to a dramatic increase of development time and cost. Therefore, there is a desire to find problems concerning integration early in the design process.

To tackle this problem, a lot of research has already been done on the usage of meta models in the development stage for ensuring consistency between subsystems. It has been found that consistency of development data could be ensured by meta model integration of requirements, class diagrams and source code [3]. A model identity card has also been proposed [4] to define a formal vocabulary that allows the interchange of data between multiple domains and actors. The concept of a Single Underlying Model (SUM) for keeping multiple views consistent [5] is an appropriate method for consistency management. Other authors claim that it is possible to achieve an increase in development speed by a factor five for cyber-physical systems by making use of meta models [6].

However, it is not clear whether and how these models can be applied to use cases from the avionics industry, for which consistency of requirements, system modelling and safety assurance are especially important. In the aerospace domain, where safety is crucial, very stringent guidelines need to be followed. Most aspects of the development process need to be validated and verified according to the ARP standards to create the required certification documents. Furthermore, certification authorities require proof that the guidelines have been followed. This can only be achieved with a consistent data set where correspondence between safety assurance, system design and requirement modeling are clearly traceable.

This paper proposes a methodology for consistency management of artefacts created throughout the system development life-cycle of an avionics system. It seeks to integrate all the artefacts in a single underlying model (SUM) for which the data is stored in a central repository. The underlying model is defined as a meta model which represents the system model, safety requirements, as well as the artefacts from safety assurance methods such as Failure Mode and Effect Analysis (FMEA) [7], Goal Structuring Notation (GSN) [8] and Functional Hazard Analysis (FHA) [2]. Furthermore, the meta model captures the associations between the meta data nodes.

The methodology mentioned above proposes the use of a single underlying model to obtain consistency between subsystems. This paper applies this method on an application in the aerospace domain, where current practices are often still making use of unrelated tools.

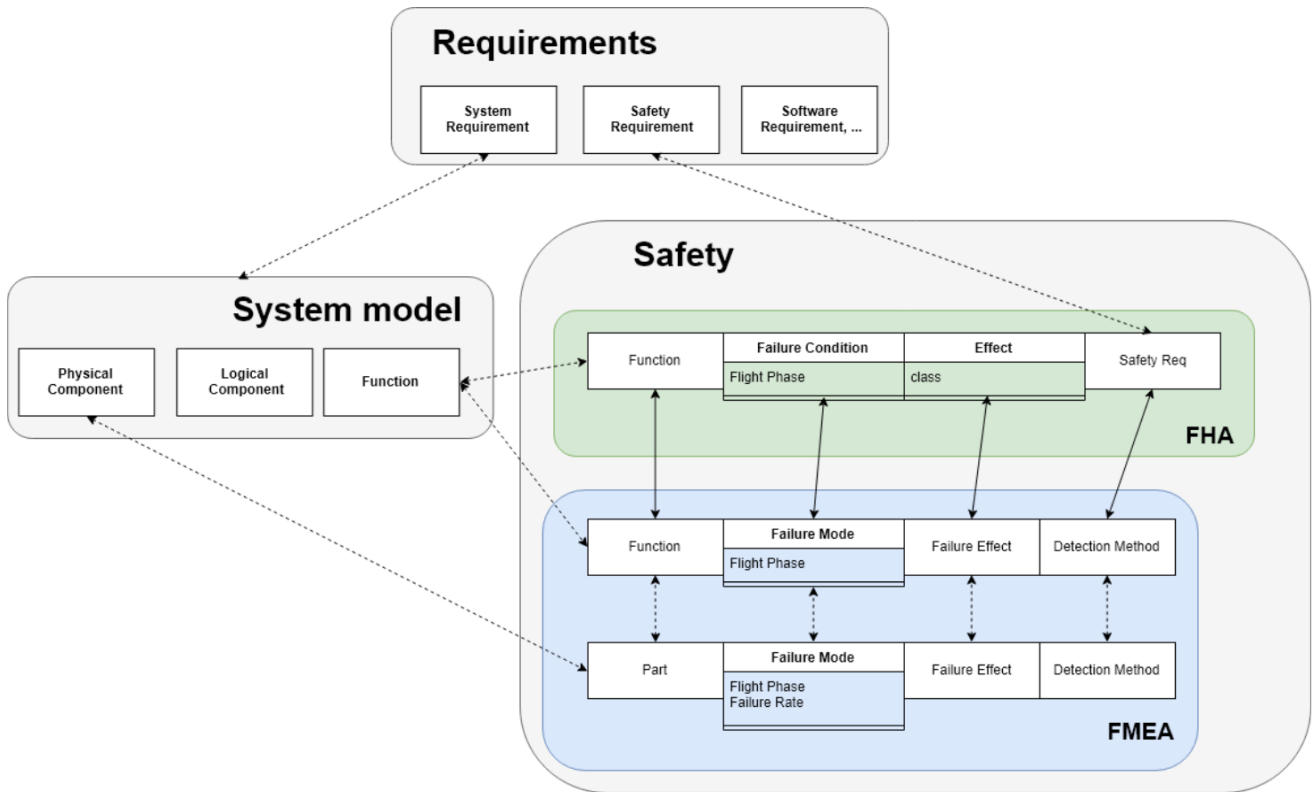

*Fig. 1: Overlap between different design artefacts*

By making use of the proposed methodology, an increase in consistency and reduction of development time can be clearly shown. Because the data is stored in one central repository, it is possible to integrate associations between the heterogeneous data artefacts that traditionally would remain in distributed silos, which leads to more consistency and traceability. Since less rework must be done during the integration phase, there will be a decrease in development time as well.

## II. METHODS

During system development, a lot of interrelated artefacts are created. All these artefacts are dependent on each other and should be kept consistent with each other. This paper proposes a methodology to help to establish and maintain traceability and consistency of these data artefacts. More specifically, requirement management, safety assurance and system modeling artefacts are considered in this paper.

### A. Development data

In the aerospace domain, safety is crucial and must be managed during the complete development lifecycle so that the system can pass certification. Safety assurance is a parallel process to system development and is closely related to other aspects of the system development lifecycle. Fig. 1 gives a simplified view on this interrelationship. For safety assurance, two types of analyses are considered: Functional Hazard Assessment (FHA) and Failure Mode and Effects Analysis (FMEA).

FHA is a bottom-up high-level approach where potential failures of high-level functions of the system are considered. Based on the expected impact of these failures, safety requirements are constructed for the system to avoid these potential failures.

FMEA is a bottom-up low-level approach where low-level functions or physical components are considered. For each of these functions/components possible failure modes and the corresponding effects are considered. If the effects are not within acceptable range, the design need to be modified together with the related (safety)requirement.

Both analyses make use of a system model which consists of physical components, logical components and functions (functions for FHA and functions/physical components for FMEA). From these safety analyses, new safety requirements are extracted. The certification documents state guidelines for the validation and verification of system development artefacts. To be able to ensure a validated data set, these certification standards require that traces are maintained between the system model components and the related safety processes. Also, the derived safety requirements need to be linked with these processes.

Both the FHA and FMEA are typically done in an isolated tool such as Excel, which proves difficulties for establishing traceability. By bringing the data together in one model, traces can be easily created, and a validated data set can be shown to the authorities.

Therefore, our method proposes to store the data from requirement specification, safety assurance and system modeling in an Application Lifecycle Management (ALM) repository such as Polarion [9]. Polarion is an extensible browser-based ALM tool for the management of requirements and other development artefacts. The repository on which it is built is version controlled (SVN) so that modifications can be easily traced back. It also has support for access permissions to allow multiple people with different access rights to work on the same project.

## FMEA Sheet

| Object | Failure Mode | Flight Phase | Mode Failure Rate | Failure Effect | FHA Effect | Detection Method |
|---|---|---|---|---|---|---|
| INES-2685 - FCS & pilot interface 2 | INES-2686 - FCS & pilot interface 2 does not provide value to computer (Taxi) | Taxi | | INES-2656 - No significant impact on the airplane | INES-2405 - No impact | INES-2402 - Compare the sensor signal 1 to the redundant sensor signal 2. if there is no val... |
| | INES-2687 - FCS & pilot interface 2 does not provide value to computer (Take off) | Take off | | INES-2401 - Elevator run away, airplane could crash | INES-2409 - Loss of plane possible | INES-2402 - Compare the sensor signal 1 to the redundant sensor signal 2. if there is no val...<br>INES-2403 - Propagate failure to cockpit |
| INES-2682 - Aircraft sensors 2 | INES-2683 - Aircraft sensors 2 do not provide values to computer (Taxi) | Taxi | | INES-2656 - No significant impact on the airplane | INES-2405 - No impact | INES-2402 - Compare the sensor signal 1 to the redundant sensor signal 2. if there is no val... |
| | INES-2684 - Aircraft sensors 2 do not provide values to computer (Take off) | Take off | | INES-2401 - Elevator run away, airplane could crash | INES-2409 - Loss of plane possible | INES-2402 - Compare the sensor signal 1 to the redundant sensor signal 2. if there is no val...<br>INES-2403 - Propagate failure to cockpit |
| INES-2679 - FCS & pilot interface 1 | INES-2680 - FCS & pilot interface 1 does not provide value to computer (Taxi) | Taxi | | INES-2656 - No significant impact on the airplane | INES-2405 - No impact | INES-2402 - Compare the sensor signal 1 to the redundant sensor signal 2. if there is no val... |
| | INES-2681 - FCS & pilot interface 1 does not provide value to computer (Take off) | Take off | | INES-2401 - Elevator run away, airplane could crash | INES-2409 - Loss of plane possible | INES-2402 - Compare the sensor signal 1 to the redundant sensor signal 2. if there is no val...<br>INES-2403 - Propagate failure to cockpit |

*Fig. 2: FMEA in Polarion*

In our work, the Polarion repository has been customized to contain next to requirements also physical components for the system model and failure modes and failure effects for the safety cases. The relationships between these components are also embedded in the Polarion project.

### B. Meta model

A meta model is constructed according to the data types (such as "function", "safety requirement", etc.) in Polarion to enable synchronization of an external visualization and editing tool with the repository. This meta model produces a formal semantics model so that the data in Polarion becomes readable for external tools. The meta model contains different data types (classes such as "requirements" and "functions"), attributes (properties such as "ID" or "description") and relations (associations such as "leads to" or "fails as") between the data types.

The meta model is constructed using Sirius [10], an eclipse-based tool for the creation of graphical modelling workbenches. Sirius makes use of Eclipse EMF [11] and its code generators to build graphical workbenches. Within the Sirius editor, links can be created between elements of the meta model (classes, attributes, relations, etc.) and customized representations (block diagram, table, flow graph, etc.). Instantiated elements from the meta model will be represented in these diagrams as defined in the Sirius editor. Sirius allows to have multiple representations (e.g. table, block diagram) for the same data (e.g. requirement set). In other words, the instantiated elements form a Single Underlying Model (SUM) from which different representations can be extracted. Modifications to these representations are pushed to the data in the SUM in Sirius, so that the changes are immediately visualized in the other representations. If, for example, a failure mode is added to a physical component in an FMEA diagram, this failure mode is also shown as a property of the physical component in a system model diagram. In parallel, changes in Sirius are also pushed to the Polarion repository so that other users working with the same repository can access the latest data.

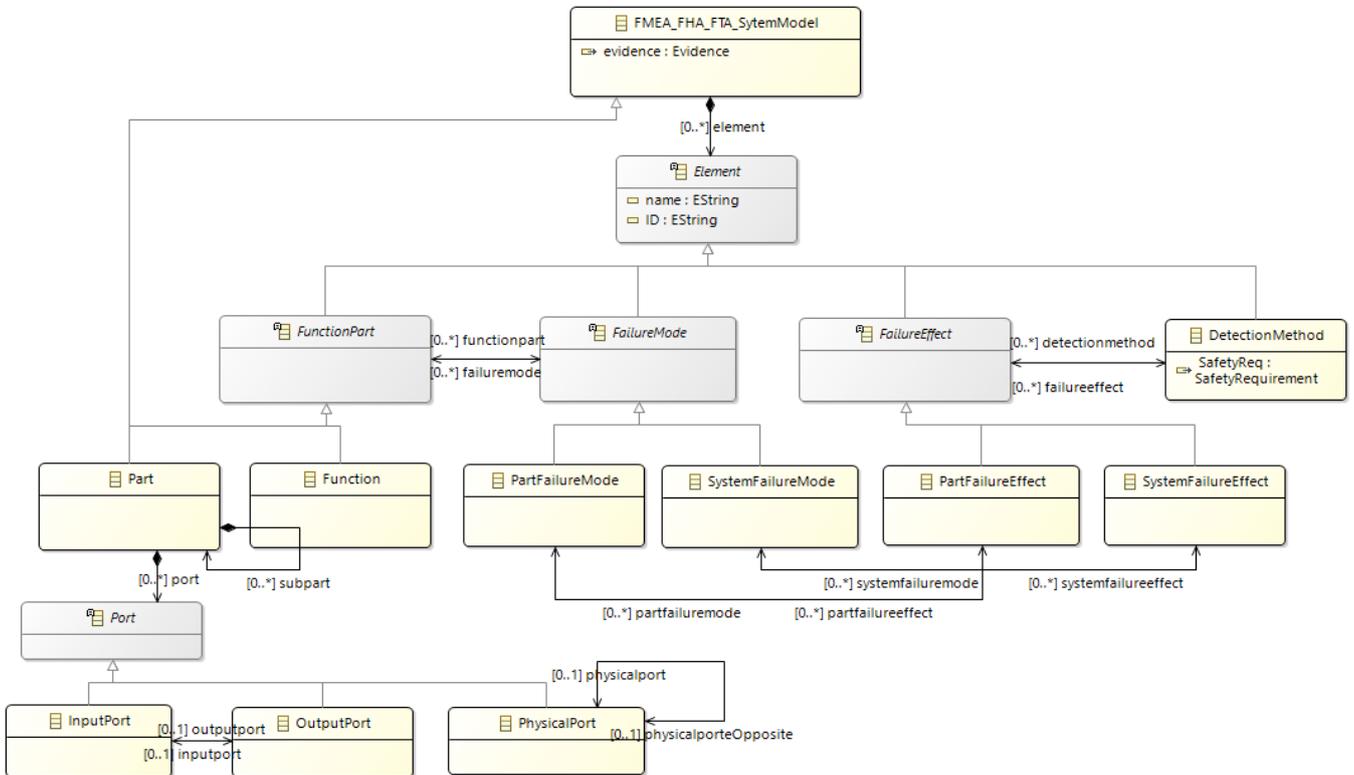

*Fig. 3: Subpart of meta model in Sirius*

Sirius gives the possibility to link elements from the meta model to custom diagrams. For this paper, custom editors have been made that refer to subsets of the meta model to represent diagrams such as a system model, FMEA or FHA.

When instances are created according to the meta model in Sirius, they inherently match both the Polarion repository and the diagrams created in Sirius. This allows to map Polarion data with its representations in Sirius.

A plugin for Sirius has been created which allows to extract and store the data in Polarion. Using this plugin, it is possible to import data from Polarion and represent in its custom Sirius diagrams. After modifications to the data in Sirius, changes are instantly sent back to the Polarion repository. This ensures that Polarion remains the central source of data which all users can interact with.

*C. Benefits*

Having a setup for development as described above can lead to an increased consistency. Because of the high certification effort in the avionics industry, consistency is key to be able to validate and verify the development artefacts as required and proposed by the standards and guidelines. The meta model imposes a structure that must be followed. It is not possible for the user to go outside of this structure, resulting in more consistent data. Data which is often contained in isolated tools such as MS Word or MS Excel is stored centrally in a shared repository. This allows inconsistencies to be found earlier in the design process.

By making use of a SUM, the development time can also be reduced. The SUM brings data artefacts from different tools together in a more efficient way with less overhead (e.g. explicitly referring to other data in other documents is not needed anymore). Because consistency is enforced, there will be less integration errors later in the design process. This will lead to less rework and thus also a reduction in development time.

In the next section, the feasibility of our approach is shown based on a case study where the requirement management, system modelling and safety assurance is performed making use of Polarion and Sirius.

### III. APPLICATION TO CASE STUDY

*A. Data repository*

In our approach, Polarion is used as a central data source. In current practice, Polarion is solely used to manage requirements and test cases. For our case study, Polarion is extended to contain, next to requirements, also system models and safety assurance cases. This is illustrated on Fig. 2 which shows an FMEA sheet in Polarion which dynamically queries mutliple workitems from its repository and represents it in a table. The first column shows the different objects (physical components) of the system. The subsequent columns represent failure modes, failure effect and detection methods for the corresponding objects. The detection methods are the result of the analysis and correspond to new safety requirements that are derived from it. These safety requirements need to be implemented so that the failure effect can be detected and corresponding actions are taken. In short, this FMEA sheet represents the interrelated nature of a problem where data that often is contained in different tools is brought together. It uses physical components from a system model, generates its own data, and the outcome is traced to safety requirements. Similar data are generated for the FHA, which is constructed based on the functions of a system model and also has safety requirements as an output.

To graphically document if the safety requirements are achieved, in the aerospace industry typically a Goal Structuring Notation (GSN) methodology is applied. It is used as the glue between safety goals (which are often in the form of safety requirements) and evidence (here FMEA).

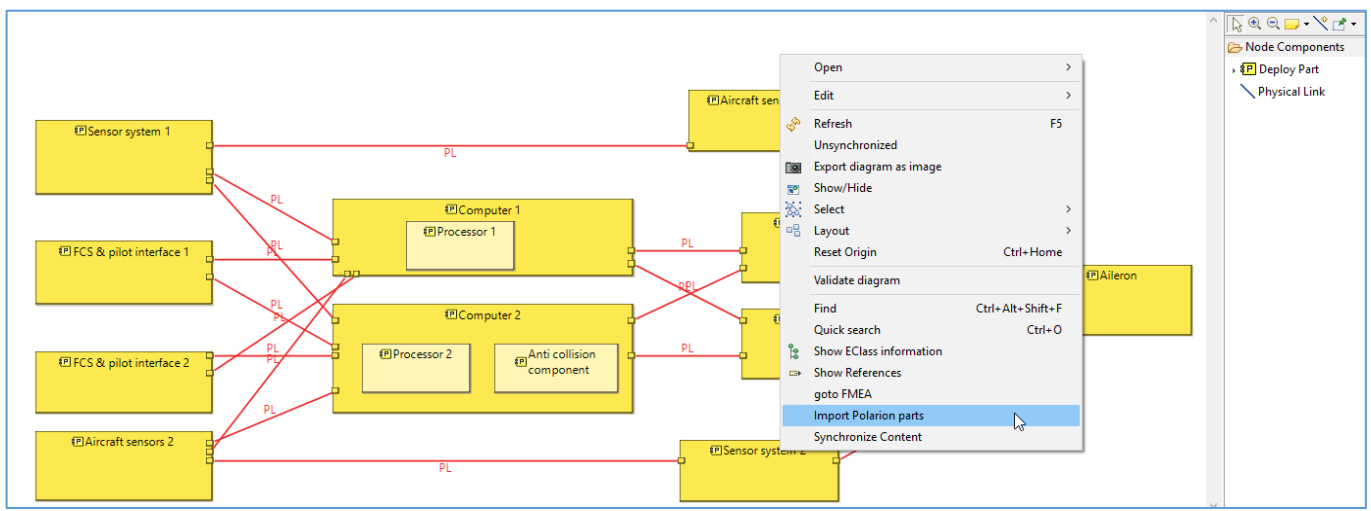

*Fig. 4: System model imported from Polarion*

It consists of following main components:

- Goal: The objective that must be achieved, often a safety requirement.
- Strategy: Describes the inference that exists between a goal and its supporting goal(s).
- Evidence: Proof of how a goal is being achieved, for instance by a safety analysis (e.g. FMEA).

The goals link to safety requirements which need to be met. A strategy links a goal and explains inference to its sub-goal(s) (e.g.: "Argumentation over all identified hazardous events"). Finally, the evidence is linked to the actual safety analysis (e.g. an FMEA) as a proof that a goal is met. For these three components custom Polarion workitems are also created so that a GSN diagram can be extracted into Sirius.

### B. Meta model mapping

The semantics of the data in Polarion is captured with a Sirius-based meta model. This meta model contains the relevant workitem types ("Part", "Failure Mode", "Failure Effect", etc) for the system model, safety assurance and safety requirements. The main attributes and interrelationships between the workitems are also defined in the meta model. Fig. 3 shows a part of the metamodel where an FMEA, FHA and system model are represented. The "SystemModel" class contains multiple Elements. These elements can be a "Function/Part", "FailureMode", "FailureEffect" or "detectionMethod". "DetectionMethod" has an association to "SafetyRequirement", which is an output of the safety assurance cases. A "Part" represents a physical component, which can have subparts, so the system can be composed of different subsystems. Similar to this figure, the rest of the meta model contains classes for the GSN and the safety requirements.

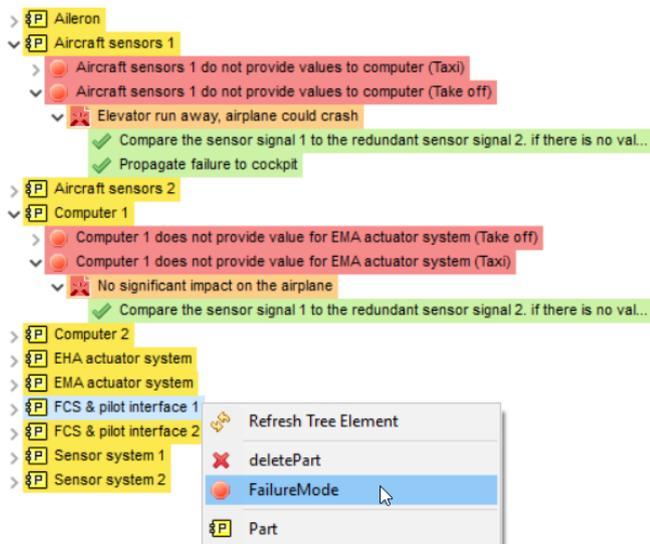

*Fig. 5: Interactive FMEA tree in Sirius*

### C. Sirius interactive visualization

Since a meta model was constructed that matches the Polarion data, it is possible to import and represent this data in specific customized diagrams. Because these diagrams are based on a SUM, multiple representations of the same data can be created. Changes to one of the representations is directly visualized in all other representations of the same data. Fig. 4 shows one of the representations where a system model of physical components and their physical connections are imported from Polarion. On the right side new parts and physical connections can be selected. Newly created components fit to the meta model and are automatically added to the Polarion repository.

The system model from Fig. 4 can be extended with safety specific information as was defined in the meta model and the Polarion project. Fig. 5 shows such an example where the physical components from the system model are extended with an FMEA. The yellow rows are the same physical components as in the system model. The red rows are the failure modes that can happen on the corresponding components. The orange rows are the failure effects of the failure mode and finally, the green rows represent the detection method that allows to mitigate the failure effect. It is possible to modify the components in this table and by doing so, changing the SUM.

Fig. 6 gives an example of a GSN that is stored in Polarion and imported in Sirius. This notation, which is typically used in the aerospace industry, can directly benefit from the SUM by tracing goals and evidence to related elements safety requirement and FMEA respectively. In the bottom of this figure there is a property, called "Fmea" that links to the corresponding safety case.

These three simplified diagrams illustrate how a single underlying model can help to integrate data which would have originally been created in isolated unrelated tools. Therefore, improved consistency and eventually shorter development time is obtained.

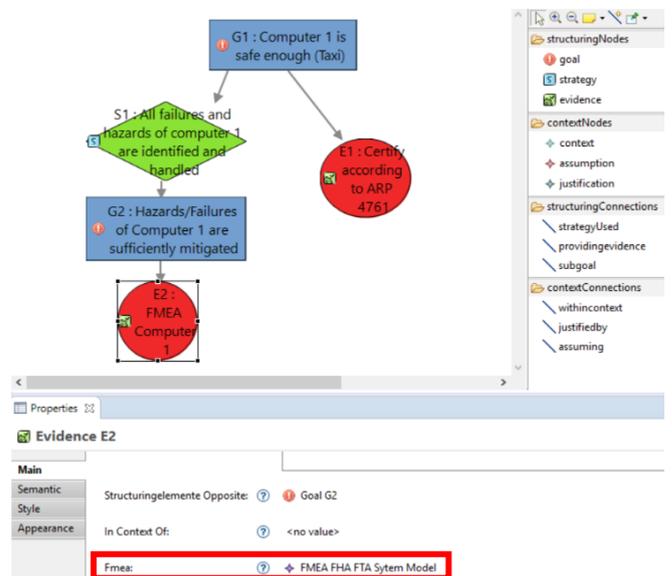

*Fig. 6: GSN diagram with goals, strategy and evidence*

## IV. LIMITATIONS

The use of a SUM, combined with a central repository can provide us with some significant improvements, however, there are also limitations to the current setup.

Firstly, if multiple people are working on the same components at the same time, only the latest changes will be stored to the latest version of the Polarion repository. There is no merge functionality to resolve possible parallel changes. Each component (requirement, physical component, failure mode, etc) is assumed to be atomic. That means that no two people should work on one component at the same time. A locking system which locks a node when it is being modified might help to avoid these problems. In any case, no data is lost when this problem occurs, the Polarion repository is based on Subversion, so all the individual changes can be retrieved if needed.

Secondly, the complete setup is based on a static meta model to allow for synchronization between Polarion and Sirius. If changes are made to this meta model after that the project has been set up, this can break the plugin that synchronizes everything, and rework needs to be done. Because of this, a lot of information on the setup of the system needs to be known beforehand so that the meta model can be constructed. This issue could be resolved by automatically extracting the meta model from the Polarion repository and generating the synchronization plugin from it. To do this, also custom editors/diagrams and the link with the generated meta model needs to be automatically constructed. This is material for future research and might be achieved by providing some extra information to the generated meta model for defining the editor type and other graphical information.

## V. Conclusion

The proposed methodology shows the applicability of an underlying meta model to an avionic system case study. In aerospace, safety and certification are crucial aspects. This paper has shown the use of a meta model throughout safety engineering activities such as FMEA, FHA and GSN integrated within a system model and linked with Safety requirements.

By moving from a document-centric to an integrated model-based approach, different interrelated data artefacts can be integrated. This paper showed the interrelationships between a system model, safety assurance and safety requirements for the avionics industry. These relationships have been captured by a meta model to allow for bringing the artefacts together, leading to a more consistent dataset and reduced development time.

A plugin has been developed to interface data from a central Polarion repository and Sirius, a graphical workbench development tool. It allows to import data from a central repository and represent it in custom Sirius diagrams. If changes were made to these diagrams, the SUM was also modified, and changes were pushed back automatically to Polarion. This ensured that Polarion contains the most recent data, allowing for better collaboration.